# The enteric nervous system is ten times stiffer than the brain

Biomechanics of the murine gut autonomic innervation

Nicolas R. Chevalier[1*], Alexis Peaucelle[2], Thomas Guilbert[3], Pierre Bourdoncle[3], Wang Xi[4]

[1]Laboratoire Matière et Systèmes Complexes, Université Paris Cité, CNRS UMR 7057, 10 rue Alice Domon et Léonie Duquet, 75013 Paris, France

[2]Institut Jean-Pierre Bourgin, INRAE, AgroParisTech, Université Paris-Saclay, 78000 Versailles, France

[3]Institut Cochin, INSERM U1016, CNRS UMR 8104, Université de Paris (UMR-S1016), 75014 Paris, France

[4]Université Paris Cité, CNRS, Institut Jacques Monod, F-75013 Paris, France
*Correspondence should be addressed to: nicolas.chevalier@u-paris.fr

## Abstract

Neural tissues of the central nervous system are among the softest and most fragile in the human body, protected from mechanical perturbation by the skull and the spine. In contrast, the enteric nervous system is embedded in a compliant, contractile tissue and subject to chronic, high-magnitude mechanical stress. Do neurons and glia of the enteric nervous system display specific mechanical properties to withstand these forces? Using nano-indentation combined with immunohistochemistry and second harmonic generation imaging of collagen, we discovered that enteric ganglia in adult mice are an order of magnitude more resistant to deformation than brain tissue. We found that glia-rich regions in ganglia have a similar stiffness to neuron-rich regions and to the surrounding smooth muscle, of ~3 kPa at 3 µm indentation depth and of ~7 kPa at 8 µm depth. Differences in the adhesion strength of the different tissue layers to the glass indenter were scarce. The collagen shell surrounding ganglia and inter-ganglionic fibers may play a key role in strengthening the enteric nervous system to resist the manifold mechanical challenges it faces.

## Significance Statement

Neural tissues of the brain are among the softest in the human body. In this study, we managed "touching" the nerve tissues of the intestine with a microscale indenter and mapped its mechanical properties. We find that the "second brain" is 10 times more rigid than central nervous system tissues and is ensheathed in a collagen cocoon which could explain this unusually high mechanical resistance.

# Main text

Unlike the central nervous system which is protected by a skull or a spine, the enteric nervous system (ENS) is embedded in a soft, deformable tissue that is constantly subject to mechanical stress. Both compressive, tensile and shear stresses act on the ENS as a result of smooth muscle contractions and deformations of the lumen by bolus. Live imaging revealed that physiological circular muscle contraction can result in up to ~50% longitudinal tensile strain on inter-ganglionic fibers[1]. Mechanical stress is also one of the essential physiological inputs mechanosensitive enteric neurons respond to[2]. Neural tissues of the brain are among the softest in the human body, with typical elastic storage modulus values in the range 0.2-1 kPa[3–5]. In contrast, smooth muscle has a resting tensile stiffness in the range 3-15 kPa[6]. Should enteric ganglia therefore be considered as soft "blobs" of brain embedded in a rigid muscular matrix? We strived here for insight into the mechanical properties of the "second brain".

Raster-scanning micro-indentation, using an atomic force microscope or equivalent setups is the gold-standard[7] to achieve the micrometric spatial resolution required to map ENS elastic properties. This method requires that the sample be flat, immobile, with the ENS directly exposed to the indenter. We addressed these challenges by peeling away the tunica muscularis of mouse duodenum, revealing regions where the ENS adhered only to the longitudinal muscle layer, with no circular smooth muscle remaining (Fig.1a). The tissue sheet was placed, in a drop of PBS, ENS facing up, on a plasma-cleaned glass slide on which polylysine had been evaporated. The PBS was carefully absorbed with tissue paper; the sample adhered within 1 min and was then rehydrated. The plasma+polylysine treatment significantly reduced the time it took for the tissue to adhere (~5 min on untreated glass), showing that it effectively mitigated artefacts that may be caused by this drying step. Post-hoc recovery of the proper range of elastic modulus values for smooth muscle further justifies this approach. Our protocol did not involve slicing the tissue[3–5], thus avoiding artefacts related to surface damage. Elastography was performed with a 51 µm diameter glass sphere attached to a cantilever (Fig.1b inset), to obtain force-displacement curves (Fig.1b) at a 5 µm resolution. Using neural-crest specific GCaMP mice, we precisely located the ENS under GFP illumination and registered it with the indenter position. Indentation depths 8.2±2.1 µm (*n*=16, ± indicates the SEM) were chosen to measure the bulk elastic properties of the ganglia (~5 µm thickness). Hertz indentation theory was used to derive the elastic modulus (Fig.1b), and due to the indentation depth being significant compared to the sample thickness, we corrected these apparent modulus values for finite-thickness effects following a reported method[8]. In total, we performed and analyzed 5948 indentations on 16 different ganglionated regions from the duodenums of 6 adult mice. Of these indentations, 27.7 % were performed on enteric ganglia, 6.7 % on inter-ganglionic fibers (IGFs) and 65.6 % on the longitudinal muscle. After indentation, samples were fixed and immunohistochemically processed to reveal cell nuclei, smooth muscle (Fig.1c), neurons and glia (Fig.1d).

In most scanned regions (14/16), there was a clear correlation between the relative-height map (topography) and ganglion position (Fig.1e), with ganglia and IGFs having significantly higher relative heights than the muscle (Fig.1h). Average elastic modulus values deduced from analysis of the whole indentation curve (8.2±2.1 µm) were 6.6±1.1, 6.5±1.0 and 6.9±1.3 kPa for ganglion, IGF and muscle respectively with no significant difference between the three tissue types (Fig.1f,i).

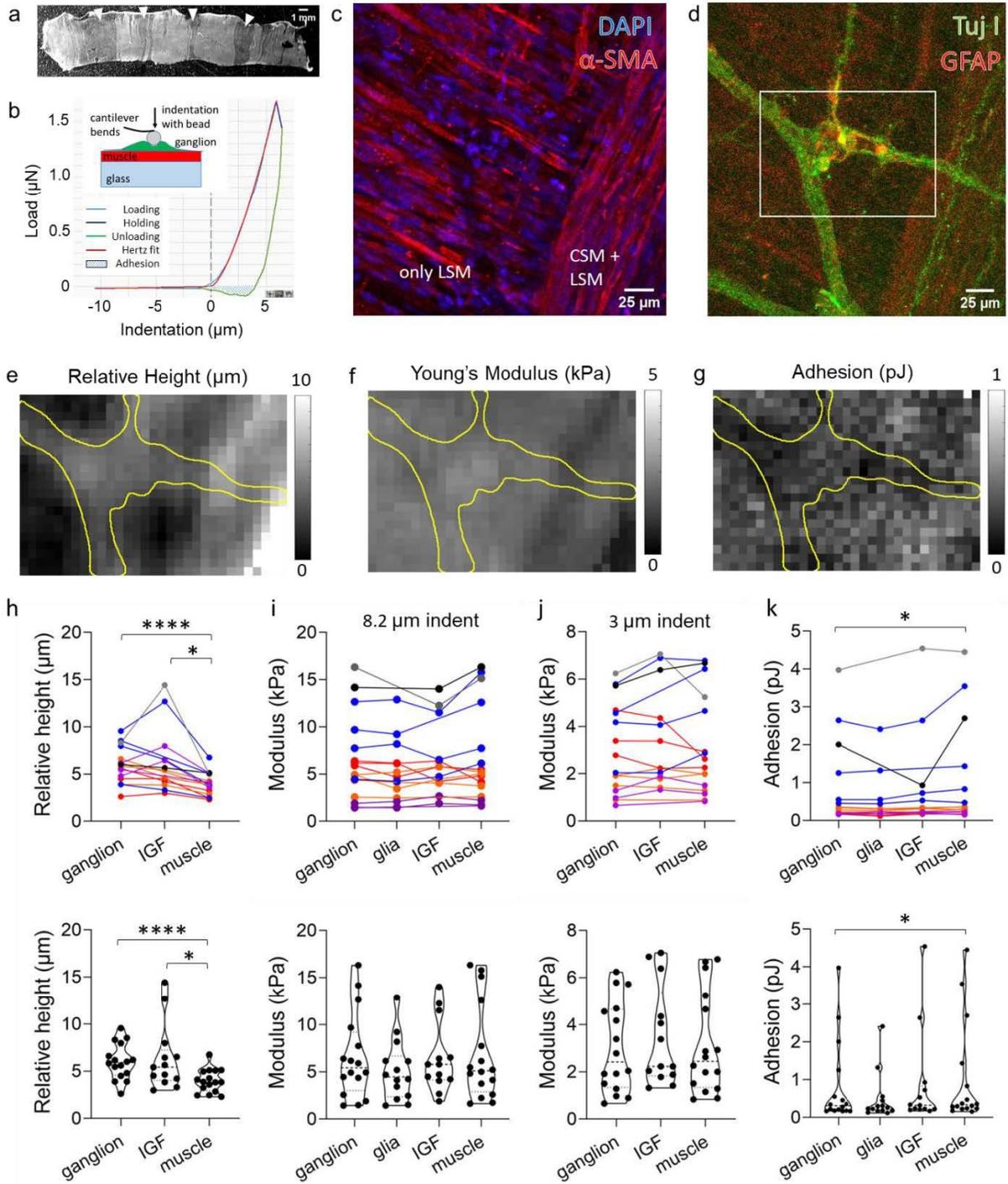

*Figure 1. The enteric nervous system has the same elasticity as the surrounding smooth muscle.* a) The peeled tunica muscularis of the mouse duodenum displayed several thinned regions (white arrowheads), where enteric ganglia were attached to the underlying longitudinal smooth muscle, but with no circular muscle on top. b) Representative load-indentation curve. Light blue: approach (loading), dark blue: holding, green: retraction (unloading), red: Hertz fit of the loading curve, dashed region: integral corresponding to the work of adhesion upon retraction. Inset: scheme of the indentation configuration. (c,d) Confocal microscopy z-max projections following immunohistochemistry. The left region of c) is devoid of circular smooth muscle (CSM), with ganglia adherent to the longitudinal smooth muscle (LSM), while the right region presents with CSM. The rectangle in d) marks the scanned

*region, with the right part of the ganglion containing most glial cells. (e,g) Maps of the relative height, Young's modulus and adhesion corresponding to the rectangular region in d). The yellow line delineates the ganglion and IGFs in d). The ganglion appears higher and slightly less adhesive than the surrounding muscular tissue, while the elastic modulus is homogeneous across the tissue. Each pixel is 5x5 µm². (h,i,j,k) Average relative height, Young's modulus at full indentation depth (8.2 µm), Young's modulus at low indentation depth (3 µm) and adhesion to the glass bead in the ganglion, glia-rich region of each ganglion, inter-ganglionic fiber (IGF) and muscle acquired from 16 scans (16 different ganglia). Each point is the average of between 20 and 370 indentations. Lines in the top row link data points that were extracted from the same map (scan). The 6 different colors indicate the 6 different mice examined. The lower row is the same data, but outlining the distribution.\*$p<0.05$, \*\*\*\*$p<0.0001$, 1-way paired ANOVA.*

When the analysis was restricted to the first 3 µm of indentation, we found elastic modulus values of 3.0±0.5, 3.5±0.6 and 3.1±0.5 kPa for ganglion, IGFs and muscle respectively (Fig.1j). Hertz fits were of excellent fidelity when restricted to only 3 µm, but this analysis is also more sensitive to extracellular material present at the ganglion surface. The elastic modulus of smooth muscle found at high indentation depths lies in the median of what is reported for this tissue[6], while the one retrieved from the small-indentation analysis is in the minimum range. The increase of elastic modulus with indentation depth (increasing strain) is consistent with other tissue indentation studies[5,9]. The scatter of elastic modulus values was due to sample-to-sample variability: elasticity was consistent across the different tissue types within the same sample and across different regions on individual samples (Fig.1i, top). Sub-analysis of glia-rich regions in individual ganglia (e.g. the right half in Fig.1d) showed no difference in mechanical properties compared to the whole ganglion (Fig.1f,i). Furthermore, IGFs, which are mostly devoid of glial cells, presented similar mechanical properties to ganglia.

The adhesion we measure by micro-indentation (Fig.1b,g,k) is that between the local tissue surface and the glass indenter: it does not yield information on cell-to-cell adhesions, which involve more specific protein complexes. We found that muscle was slightly (20%) but significantly more adhesive than enteric ganglia (Fig.1g,k). Glia-rich regions in ganglia showed no difference of adhesive properties compared to the whole ganglion (Fig.1g,k).

Although glial (γλία, greek for glue) cells are often ascribed a "structural" or "supporting" role in ENS ganglia[10], our findings show that glia-rich regions did not display a particular mechanical resistance, or adhesion to the glass surface of the indenter. Previous biomechanical investigations on neural and glial cells isolated from the hippocampus and retina[3] found that glia are actually ~1.5-2 times softer than neurons. Instead of acting as mechanical stiffeners of the ganglion, we hypothesize that glia could instead play a structural role in plastically accommodating neuronal volume changes, as observed in the retina[11].

Enteric ganglia are surrounded by a thin shell of collagen I, first revealed by electron microscopy[12], which we evidence here using a label-free optical technique – second harmonic generation (SHG). The collagen shell was visible on whole gut vibratome sections (Fig.2a-c), and was firmly attached to the ganglion and IGFs, as it was still present after stripping the overlying circular smooth muscle layer (Fig.2.d). This collagen shell likely contributes to ENS mechanical integrity, and may explain its high stiffness compared to brain tissue.

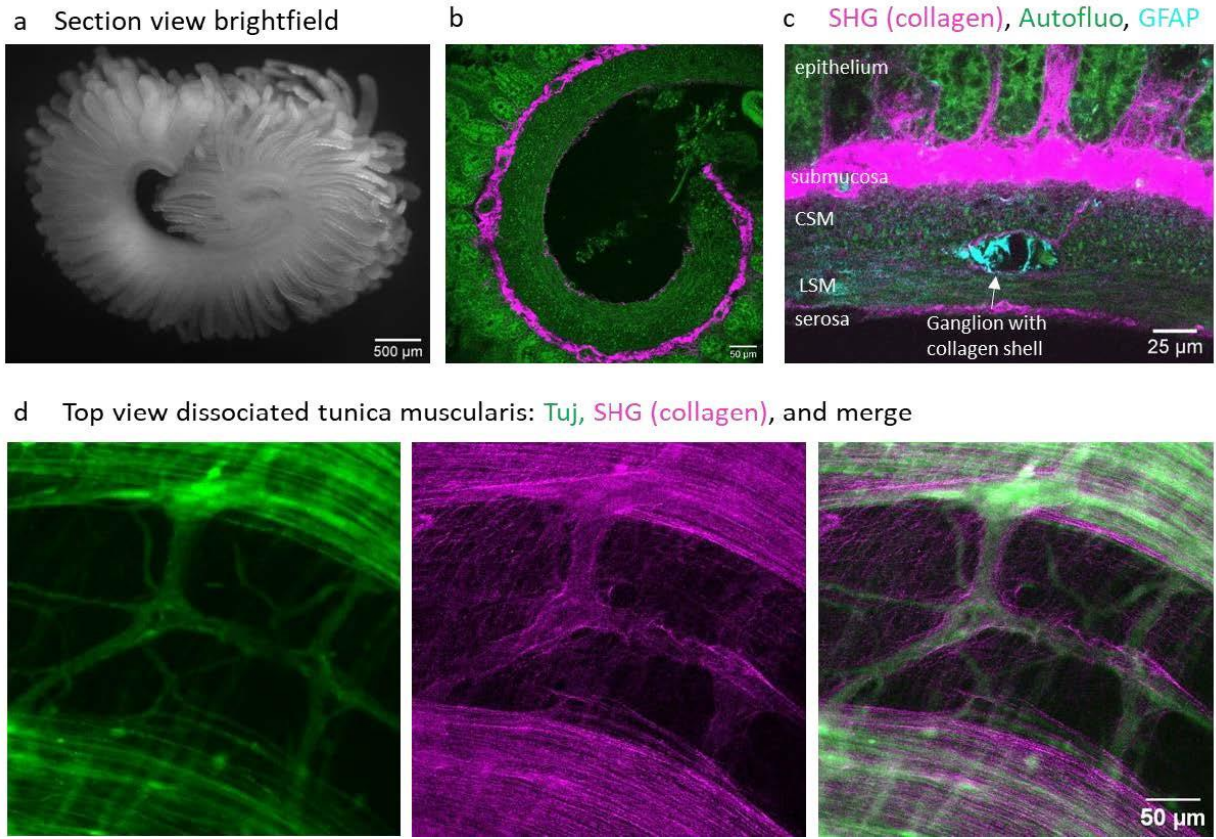

*Figure 2. A fibrillar collagen shell is firmly anchored to the myenteric nerve plexus.* (a) Brightfield picture of duodenum transverse vibratome slice, one duodenum wall was cut causing the sample to wrap on itself, mucosa at the periphery, tunica in the center. (b) SHG showing the dominant collagen I signal from the submucosa. (c) At higher laser intensity, SHG reveals a collagen shell surrounding a myenteric ganglion and lining the smooth muscle layers. (d) Tuj1 IHC staining (green) and SHG (magenta) of the dissociated tunica muscularis preparation used for the indentation test shows that the collagen is firmly attached to the ganglion and inter-ganglionic fibers. This was confirmed for all ganglia (n=4/4) imaged in the preparation. Non-specific Tuj1 staining also labeled here the smooth muscle layers.

In conclusion, we measured for the first time the elastic and adhesive properties of enteric ganglia, of inter-ganglionic fibers and of the surrounding smooth muscle. We find that this ensemble behaves mechanically as a homogeneous structure with a Young's modulus of ~6.5 kPa at high indentation depth (8 µm) and of ~3 kPa at low indentation depth (3 µm). Mouse or rat brain tissue have an elastic modulus in the range 0.1-2 kPa[7] (Table 1). The study of Christ et al.[5] in particular is methodologically almost identical to ours and finds brain tissue moduli of 0.3-0.45 kPa. Our findings therefore show that mouse ENS ganglia are about an order of magnitude stiffer than central nervous system brain tissue. This high stiffness could be due to the collagen I shell surrounding ganglia and inter-ganglionic fibers, to the cell-specific mechanical properties of enteric neurons and glia, or both. The second brain is much more resistant to mechanical stress than the central nervous system, making it better suited to withstand the innumerable mechanical stresses our gut is subject to.

| Study | Sample | Analysis Method | Bead diameter (µm) | Indentation depth (µm) | Modulus (kPa) |
|---|---|---|---|---|---|
| Elkin et al.[4] | P8-10 rat hippocampus | Hertz fit | 25 | 3 | 0.1-0.3 |
| Elkin et al.[9] | Adult rat cerebellum | Hertz fit + non-linear elasticity at 30% strain | 25 | ~1.5 | 1 (cortex) 0.4-2 (hippocampus) |
| Christ et al.[5] | Adult rat cerebellum | Hertz fit | 40 | 3 | 0.3 (white matter) 0.45 (grey matter) |
| Menal et al.[13] | Adult mouse cortex | Hertz fit | 25 | 1.5 | 0.6 |
| This study | Adult enteric mouse ganglia | Hertz fit | 51 | 3 8 | 3 6.6 |

Table 1. Comparison of elastic modulus found here with similar indentation studies in the rat or mouse brain.

## Materials & Methods

**Ethics and animal samples.** Mice were hosted at the Institut Jacques Monod husbandry. Tissue samples were collected from 2 to 6 month old female HTPA:GCaMP mice (7 in total) on a C57BL/6 background. The HTPA:GCaMP mouse was previously described and expresses GCaMP6f specifically in neural crest cells and their neural and glial derivatives. Mice were killed by cervical elongation and the gastrointestinal tract dissected in PBS. The method used to kill the mice conforms to the guidelines of CNRS and INSERM animal welfare committees. Killing of mice for retrieval of their organs is a terminal procedure for which neither CNRS nor INSERM assign ethics approval codes and hence none are given here.

**Sample preparation.** A 4 mm diameter aluminum rod was inserted in the duodenum lumen. We incised the tunica muscularis along the whole segment length with sharp tweezers, and peeled it from the intestine with a humidified cotton swab[14]. The tissue sheet presented with numerous regions in which the ENS was exposed (Fig.1a) and was adhered to a glass slide as described in the main text.

**Indentation.** Elastography was performed with the Chiaro nano-indenter (Optics11, the Netherlands) mounted on an inverted epifluorescence microscope (Olympus, IX83P2ZF). The OP1550 interferometer to precisely detect cantilever bending[15]. All measurements were performed at room temperature in 3 mL of PBS with 1 mM $Ca^{2+}$ and 0.5 mM $Mg^{2+}$, with the same probe consisting of a d=51 µm diameter glass sphere attached to a 0.47 N/m cantilever, calibrated on the same day prior to each experiment. This cantilever stiffness was chosen because softer ones (0.015 N/m) would not allow to indent beyond ~5 µm, and would at times bend too much during retraction due to tissue adhesion, causing the scan to become unstable. A region of the sample where the ENS was exposed was first located, the probe was positioned close to an enteric ganglion, and its initial position relative to the ganglion recorded in GFP, which captured the distinct ENS GCaMP signal. Maps ranged from 21x21 pixel to 21x31 pixels, with a step size of 5 µm, that was increased to 7.5-12.5 µm for bigger ganglia (n=5/16 maps). Indentation speed was 20 µm/sec, with a fixed z-displacement of 20 µm from a reference z-position 5 – 10 µm above the tissue, resulting in sample indentation depths δ in the range of 5-12 µm (average 8.2±2.1 µm); Hertz indentation theory assumptions were met for small indentation analysis (δ=3 µm<<R), but only partially so for high indentation (δ=8.2 µm<R). The probe was then retracted by 60 µm to warrant detachment from the sample. Recording a map took ~1h30. Up to 4 maps (4 ganglia) were collected from the same sample. We did not observe any degradation of sample elastic properties during a recording session.

**Immunohistochemistry (IHC).** The sample was fixed immediately after elastography in 4% PFA for 5 min. Samples were permeated in 1% BSA and 0.1% triton in PBS for 15 min, then incubated in 1:500 primary GFAP rabbit antibody (Agilent Z033429-2) for 1 h, washed (3x), next incubated in 1:1000 DAPI, 1:500 conjugated βIII-tubulin (Tuj1) AF488 (sc-80005), 1:1000 conjugated α-SMA Cy3 (Sigma C6198) and in 1:500 anti-rabbit AF647 secondary antibody (Thermofisher) for 1 h, and washed (2x). The sample was imaged on the same day with a Zeiss LSM780 upright confocal microscope, using the elastography GFP images to locate the indented ganglia. 1 µm z-stacks were recorded at x20 magnification.

**Data analysis.** Cantilever force-distance curves (Fig.1b) were fitted with the Hertz model using the DataViewer software (Optics11) to extract the apparent Young's modulus $E_{app}$ and local relative tissue height $h_{rel}$ (topography). The lowest point of each scan was defined as $h_{rel} = 0$. The thickness of the longitudinal muscle layer was measured by confocal microscopy to be $e \approx 20$ µm. For each point of the elastography map, the local tissue thickness was input as $h = e + h_{rel}$, and correction for finite thickness effects to the Young's modulus E calculated following formula (6, 7a, 7b) of Long et al.[8]: $E = E_{app} \frac{1+2.3\omega}{1+1.15\omega^{1/3}+\alpha\left(\frac{R}{h}\right)\omega+\beta\left(\frac{R}{h}\right)\omega^2}$ with $\omega = (\frac{R\delta}{h^2})^{3/2}$, $\alpha\left(\frac{R}{h}\right) = 10.05 - 0.63\sqrt{\frac{h}{R}}\left(3.1 + \frac{h^2}{R^2}\right)$ and $\beta\left(\frac{R}{h}\right) = 4.8 - 4.23h^2/R^2$, where $R$ is the indenter radius and $\delta$ is the local indentation depth. Calculated moduli only differed by ~5% between the slip and no-slip (of the indenter on the tissue) assumptions, so we only show results for the slip case. E was typically a factor ~2 lower than $E_{app}$ for indentation depths 8.2 µm; the correction was more important for the muscle (low thickness region) than for the ganglia (higher thickness regions). Adhesion work A (in picoJoules, pJ) was computed as the integral of the retraction curve that extended the cantilever (see Fig.1b). h, E and A grey-scale maps were then assembled in a stack together with the GFP images of the ganglion. The image was divided in different ROIs corresponding to ganglion (1 per scan), IGF (0 to 4 per scan), and muscle (2 to 4 per scan), and the area-weighted average values of h, E and A measured over the different regions. IHC images were used to locate glia-rich regions.

**Second harmonic generation microscopy (SHG).** SHG is an exquisite method to image collagen I fibers in the gut[16,17]; it is much more resolved than collagen I IHC because, being a label-less technic, it is not hampered by non-specific signal. Multiphoton imaging was performed on 1 mm thick transverse vibratome sections of PFA-fixed wildtype adult mouse duodenum embedded in agarose (Fig.2a-c), and on PFA-fixed dissociated tunica muscularis preparations (Fig.2d) with an upright Leica SP8 DIVE (Wetzlar, Germany) coupled with a pulsed laser controlled by LAX software. SHG (ex 950 nm, em 465-485nm), two-photon autofluorescence (ex 950 nm, em 500-550 nm), AF488 (ex 488 nm, em 500-600 nm), AF647 (ex 638 nm, em 691-728nm) were acquired with a 25x objective immersed in PBS.

## Statements & acknowledgments

**Corresponding Author ORCID**: 0000-0002-9713-1511

**Conflict of interest statement**: The authors declare that they have no conflict of interest.

**Author contributions:** NRC led the project, performed experiments, analyzed data, wrote and revised the paper. PB performed experiments. AP, TG and WX performed experiments and revised the paper.

**Data availability statement**: All of the analyzed data is presented in the manuscript and its supplementary information. Raw data is available from the authors upon request.

**Grant support**: This research was funded by the Agence Nationale de la Recherche ANR GASTROMOVE-ANR-19-CE30-0016-01, by the Labex "Who AM I ?" ANR-11-LABX-0071, and by the Imaging platform BioEmergences-IBiSA, ANR-10-INBS-04 and ANR-11-EQPX-0029.

**Acknowledgments:** We are grateful to the whole animal husbandry staff at Institut Jacques Monod, and to Sylvie Dufour for providing the Ht-PA::Cre mouse line We thank Meenakshi Rao, Amy Shepherd and Anoohya Muppirala for contributing samples, and Vincent Fleury for stimulating discussions. We are grateful to the Institut Cochin Imagic Platform for SHG imaging.

## References

1. Chevalier, N. R., Agbesi, R. J. A., Ammouche, Y. & Dufour, S. How Smooth Muscle Contractions Shape the Developing Enteric Nervous System. *Front. Cell Dev. Biol.* **9**, 1402 (2021).

2. Spencer, N. J., Dinning, P. G., Brookes, S. J. & Costa, M. Insights into the mechanisms underlying colonic motor patterns. *J. Physiol.* **594**, 4099–4116 (2016).

3. Lu, Y. B. *et al.* Viscoelastic properties of individual glial cells and neurons in the CNS. *Proc. Natl. Acad. Sci. U. S. A.* **103**, 17759–17764 (2006).

4. Elkin, B. S., Azeloglu, E. U., Costa, K. D. & Morrison, B. Mechanical heterogeneity of the rat hippocampus measured by atomic force microscope indentation. *J. Neurotrauma* **24**, 812–822 (2007).

5. Christ, A. F. *et al.* Mechanical difference between white and gray matter in the rat cerebellum measured by scanning force microscopy. *J. Biomech.* **43**, 2986–2992 (2010).

6. Matsumoto, T. & Nagayama, K. Tensile properties of vascular smooth muscle cells: Bridging vascular and cellular biomechanics. *J. Biomech.* **45**, 745–755 (2012).

7. Babu, P. K. V. & Radmacher, M. Mechanics of brain tissues studied by atomic force microscopy: A perspective. *Front. Neurosci.* **13**, 1–9 (2019).

8. Long, R., Hall, M. S., Wu, M. & Hui, C. Y. Effects of Gel Thickness on Microscopic Indentation Measurements of Gel Modulus. *Biophys. J.* **101**, 643 (2011).

9. Elkin, B. S., Ilankovan, A. & Morrison, B. Age-dependent regional mechanical properties of the rat hippocampus and cortex. *J. Biomech. Eng.* **132**, 1–10 (2010).

10. Gulbransen, B. D. Enteric Glia. *Colloq. Ser. Neuroglia Biol. Med. from Physiol. to Dis.* **1**, 1–70 (2014).

11. Uckermann, O. *et al.* Glutamate-evoked alterations of glial and neuronal cell morphology in the guinea pig retina. *J. Neurosci.* **24**, 10149–10158 (2004).

12. Gabella, G. Fine structure of the myenteric plexus in the guinea-pig ileum. *J. Anat.* **111**, 69 (1972).

13. Menal, M. J. *et al.* Alzheimer's disease mutant mice exhibit reduced brain tissue stiffness compared to wild-type mice in both normoxia and following intermittent hypoxia mimicking sleep apnea. *Front. Neurol.* **9**, (2018).

14. Huang, Z. *et al.* An efficient approach for wholemount preparation of the myenteric plexus of rat colon. *J. Neurosci. Methods* **348**, 109012 (2021).

15. Van Hoorn, H., Kurniawan, N. A., Koenderink, G. H. & Iannuzzi, D. Local dynamic mechanical analysis for heterogeneous soft matter using ferrule-top indentation. *Soft Matter* **12**, 3066–3073 (2016).

16. Chevalier, N. R. *et al.* A neural crest cell isotropic-to-nematic phase transition in the developing mammalian gut. *Commun. Biol.* **4**, 770 (2021).

17. Chevalier, N. R. *et al.* How tissue mechanical properties affect enteric neural crest cell migration. *Sci. Rep.* **6**, 20927 (2016).